\documentclass[aps,prb, amsmath,amssymb, twocolumn, showpacs,longbibliography]{revtex4-1}

\usepackage{graphicx} 
\usepackage{dcolumn}
\usepackage{bm}
\usepackage{amsmath}
\usepackage{amssymb}
\usepackage{color}
\usepackage{bbold}
\usepackage{bm}
\usepackage{mwe}

\def \e{{\epsilon}}

\renewcommand{\vec}[1]{{\boldsymbol #1}}

\newcommand{\bs}[1]{\boldsymbol{#1}}
\newcommand{\ket}[1]{|#1\rangle}
\newcommand{\bra}[1]{\langle#1|}
\newcommand{\ave}[1]{\langle #1 \rangle}

\newcommand{\llangle}[1][]{\savebox{\@brx}{\(\m@th{#1\langle}\)}%
  \mathopen{\copy\@brx\kern-0.5\wd\@brx\usebox{\@brx}}}
\newcommand{\rrangle}[1][]{\savebox{\@brx}{\(\m@th{#1\rangle}\)}%
  \mathclose{\copy\@brx\kern-0.5\wd\@brx\usebox{\@brx}}}

\begin{document}
\title{Activating many-body localization in solids by driving with light} 
\author{Zala Lenar\v{c}i\v{c}$^1$}
\author{Ehud Altman$^2$}
\author{Achim Rosch$^1$}
\affiliation{$^1$Institute for Theoretical Physics, University of Cologne, D-50937 Cologne, Germany\\
$^2$Department of Physics, University of California, Berkeley, California 94720, USA}

\begin{abstract}
Due to the presence of phonons, many body localization (MBL) does not occur in disordered solids, even if disorder is strong. Local conservation laws characterizing an underlying MBL phase decay due to the coupling to phonons. Here we show that this decay can be compensated when the system is driven out of equilibrium. The resulting  variations of the local temperature provide characteristic fingerprints of  an underlying MBL phase. We consider a one-dimensional disordered spin-chain which is weakly coupled to a phonon bath and weakly irradiated by white light. The irradiation has weak effects in the ergodic phase. However, if the system is in the MBL phase irradiation induces strong temperature variations of order 1 despite the coupling to phonons. Temperature variations can be used similar to an order parameter to detect MBL phases, the phase transition and an MBL correlation length.
\end{abstract}

\maketitle


A quantum many body system subjected to strong disorder can be many-body localized and thus fail to thermalize when evolving under its own dynamics \cite{Basko06,Mirlin05}. This phenomenon has attracted a lot of interest as an example of a novel dynamical state of matter. In the case of a fully MBL state, where all the many-body eigenstates of the hamiltonian are localized, the system is characterized by an extensive set of local integrals of motion \cite{Vosk13,Serbyn13-1,Huse13,ros15}. 

The local conservation laws persist without fine-tuning, which makes MBL more robust than conventional integrability. Like integrable models, however, many-body localization cannot survive even the weakest static coupling of the system to an external bath of delocalized excitations  \cite{nandkishore17}. 
Any such coupling would lead to thermalization, 
therefore all direct experimental demonstrations of MBL were so far achieved with ultra cold atomic systems \cite{Bloch15} 
as well as trapped ions \cite{Monroe16}, which can be extremely well isolated from the environment. In solids, by contrast, the electronic degrees of freedom are inevitably coupled to phonons and the ensuing thermal state shows no sign of the local integrals of motion.

\begin{figure}[h!]
\center
\hspace{.085\linewidth}
\includegraphics[width=.635\linewidth]{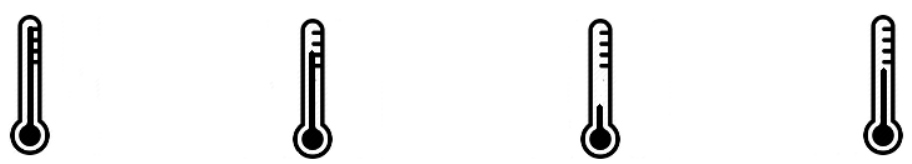}\\
\vspace{-.05\linewidth}
\includegraphics[width=.87\linewidth]{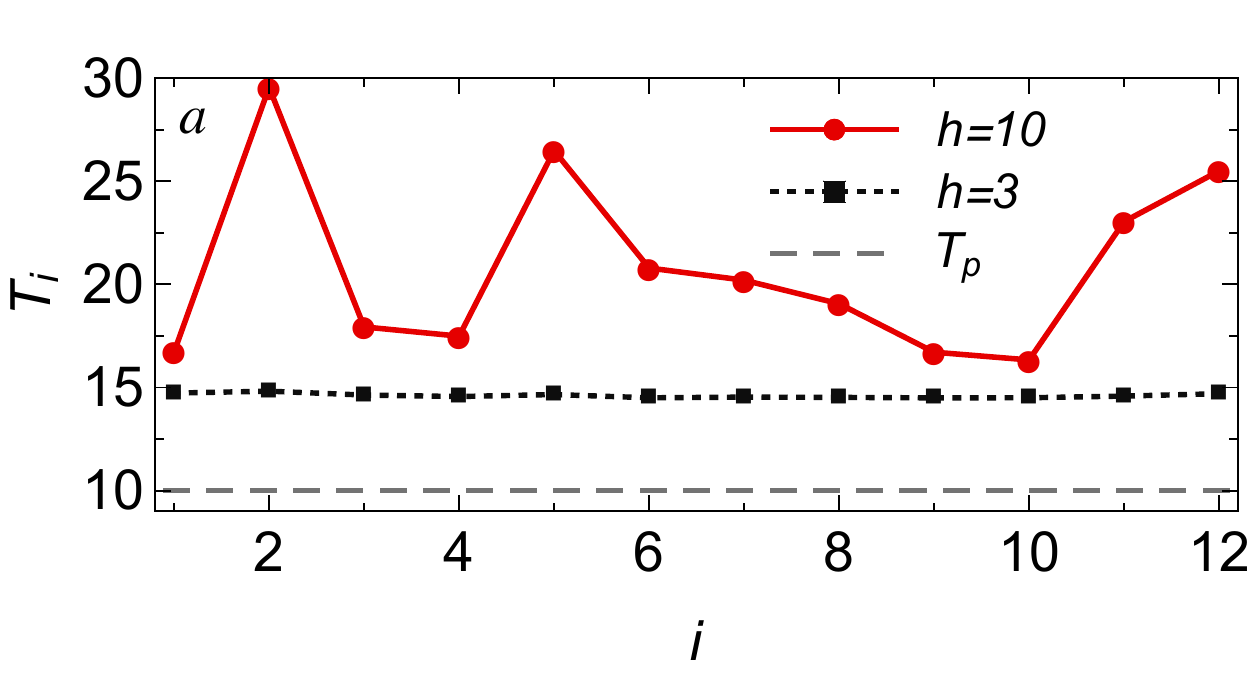}\\
\hspace{.027\linewidth}
\includegraphics[width=.813\linewidth]{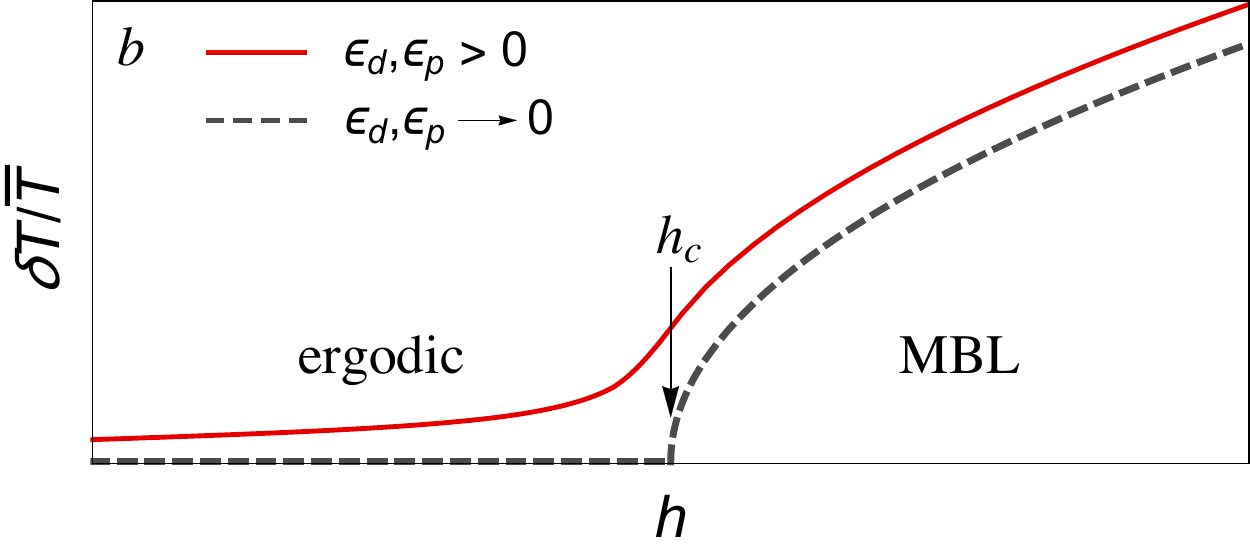}
\caption{\label{FigExampleLocTemp} Upper panel: profile of local temperatures in systems coupled to a phonon bath at temperature $T_p$ and driven weakly by white light. The temperature variations are large in the MBL phase (red circles), while they are vanishing in the ergodic phase (black squares). Lower panel: a schematic depiction of the standard deviation of local temperatures versus disorder strength. A sharp phase transition is expected in the limit of vanishing coupling to the phonons and the drive, i.e. $\epsilon_d,\epsilon_p\to 0$, while $\epsilon_d/\epsilon_p\to \text{const}$.}
\end{figure}

In this paper we argue that the local integrals of motion of an electronic system can be "reactivated" by driving the system to a non-equilibrium steady state.  
In essence, the driving counters the relaxation with the phonon bath, giving rise to a new steady state in which the value of the local integrals of motion is set by a local balance between the phonons and the drive. In the limit of weak drive and weak coupling to phonons this scheme allows to make a sharp distinction between the steady states obtained with the dominant Hamiltonian in the MBL phase compared to an ergodic one, as demonstrated in Fig.\ref{FigExampleLocTemp}.

In previous work, we developed a formalism for computing the steady state density matrix of integrable systems subject to weak driving and coupling to baths \cite{lange17,lenarcic18,lange18}, which is also applicable here. In the limit of weak driving steady state expectation values can be approximately computed using a generalized Gibbs ensemble adjusted to expectation values of the integrals of motion, determined by rate equations. Fig. \ref{FigExampleLocTemp} shows how this scheme plays out in the MBL phase compared to a conventional thermalizing phase. In the fully MBL system, there is always a set of integrals of motion related to the local energy density. Hence driving the system gives rise to widely varying local temperatures. In an ergodic system, on the other hand, only the global energy is conserved, hence when it is weakly driven the system equilibrates to a thermal state characterized by a single temperature.

In a similar setup with a disordered system weakly coupled to a bath and a monochromatic drive Refs. \onlinecite{deLuca15,deLuca16b} studied optimization of nuclear polarization. Nuclear polarization is optimized when the underlying system is close to the localization transition. In agreement with our results, they observe that an equilibrium description in terms of spin temperature can be applied only in the ergodic phase.
Also Ref.~\onlinecite{Basko07} has previously proposed a different approach to detect indirect signatures of many-body localization in electron systems. The emphasis of that work is on signatures of a finite temperature localization transition that persist despite the broadening of the transition due to coupling to phonons. 
Recently possible indications of the proximity to such a transition were seen in a disordered InO film \cite{Shahar}. These effects are, however indirect, and may not be unique for MBL. The effects we discuss in this paper, by contrast, provide a direct unambiguous signature of MBL.

Our goal is to describe a strongly disorded, interacting electron system in a solid weakly coupled to phonons and irridiated by light, $H=H_f+H^0_{p}+H_{fp}+H_d$.
To  simplify the numerical analysis, we consider a one-dimensional model of spinless fermions with periodic boundary conditions at half-filling
\begin{equation}
H_f=\tilde t \sum_{i=1}^{N}(c_i^\dagger c_{i+1} + c_{i+1}^\dagger c_i) + V_i \, n_i + U n_i n_{i+1}
\end{equation}
which is related to the Heisenberg model of spins via Jordan-Wigner transformation. We use interaction strength $U=2$ corresponding to the isotropic point of the Heisenberg model. The lattice constant $a$ is set to $a=1$ as well as $\tilde t=1$.
The random local potential $ V_i $ is drawn from a box-distribution, $V_i\in [-h,h]$. This model and its variants \cite{oganesyan07,pal10,luitz15,geraedts17} 
have been studied extensively and are known to show an (infinite temperature) MBL transition at a critical disorder strength of about $h \approx 7$, see e.g. \cite{pal10,luitz15}. 

The fermions interact with three dimensional acoustic phonons $H^0_{p}=\sum \omega_{\bs q} a^\dagger_{\bs q} a_{\bs q}$ with dispersion $\omega_{\bs q}=v|\bs{q}|, v=\tilde t$ which couple to the electrons through the hopping matrix element
\begin{align}
H_{fp}&= \epsilon_p 
\sum_{q_x} \int \frac{d q_\perp^2}{(2 \pi)^2}  \, (a_{\bs q} + a_{-\bs q}^\dagger) \frac{i q_x}{\sqrt{2\omega_{\bs q}}} \ H_{q_x}, \label{EqHph}\\
H_{q_x}&=\frac{1}{\sqrt{N}} \sum_{j}\tilde{t} e^{i q_x j}(c_{j+1}^\dagger c_j + c_{j}^\dagger c_{j+1})\notag
\end{align}
The dimensionless parameter $\epsilon_p$ controls the strength of electron phonon interaction. Due to periodic boundary conditions in $x$, the (dimensionless) momenta take quantized values $q_x=\frac{2 \pi}{N}n_x$ while the perpendicular momenta are continuous. The three dimensional phonons act as a thermal bath with a fixed temperature $T_{p}$.
 
At the same time the system is driven out of equilibrium due to irradiation by white light arising, e.g., from a light bulb with a very high temperature $T_d \gg \tilde{t},h,T_p$. 
The light couples to the current operator
 \begin{align}
H_{d}&= \epsilon_d A(t) \sum_{i}i \tilde{t} \, (c_{i+1}^\dagger c_i   - c_{i}^\dagger c_{i+1}) \label{EqHd}
\end{align}
Here the dimensionless vector potential is given by $\epsilon_d A(t)$. We assume that $A(t)$ is a delta-correlated classical field $\langle A(t) A(t') \rangle = 2\pi \delta(t-t')$ and $\epsilon_d$ parametrizes the amplitude of the electric fields. 

We calculate the steady-state of the driven interacting system in the limit of weak coupling to phonons and light, 
$\lim_{\epsilon_p,\epsilon_d \to 0} \lim_{t\to\infty}\rho(t)= \sum_n p_n |n\rangle \langle n|$. In this limit the density matrix is diagonal in terms of eigenstates of $H_f$ with amplitudes $p_n$ determined from rate equations with transition rates due to coupling to phonons and driving determined from Fermi's golden rule, see Methods.

Properties of obtained steady state are studied through the behavior of local temperatures at different sites.  To define local temperatures out of equilibrium, we directly model a 'thermometer' by infinitesimally coupling a phonon bath with temperature $T_j$ to the tunneling term $c^\dagger_j c_{j+1} +c^\dagger_{j+1} c_{j}$. The temperature $T_j$ is determined from the condition that the energy current to the thermometer vanishes, see Methods for details. Note that out of equilibrium the precise value of $T_j$ depends on the type of thermometer one is using. However, the qualitative difference in behavior between the ergodic  and  MBL phases is not sensitive to such details. Experimentally, there are various methods to measure local temperatures, including the measurement of thermoreflectance \cite{cahill14}, scanning thermal microscopy \cite{cui17}, fluorescent microthermal imaging \cite{barton96}. Finally, scanning light sources \cite{sternbach17} for local Raman spectroscopy \cite{anderson00} can be used to obtain local temperatures by comparing Stokes- and anti-Stokes lines of suitable transitions.

\begin{figure}[t]
\center
\includegraphics[width=.87\linewidth]{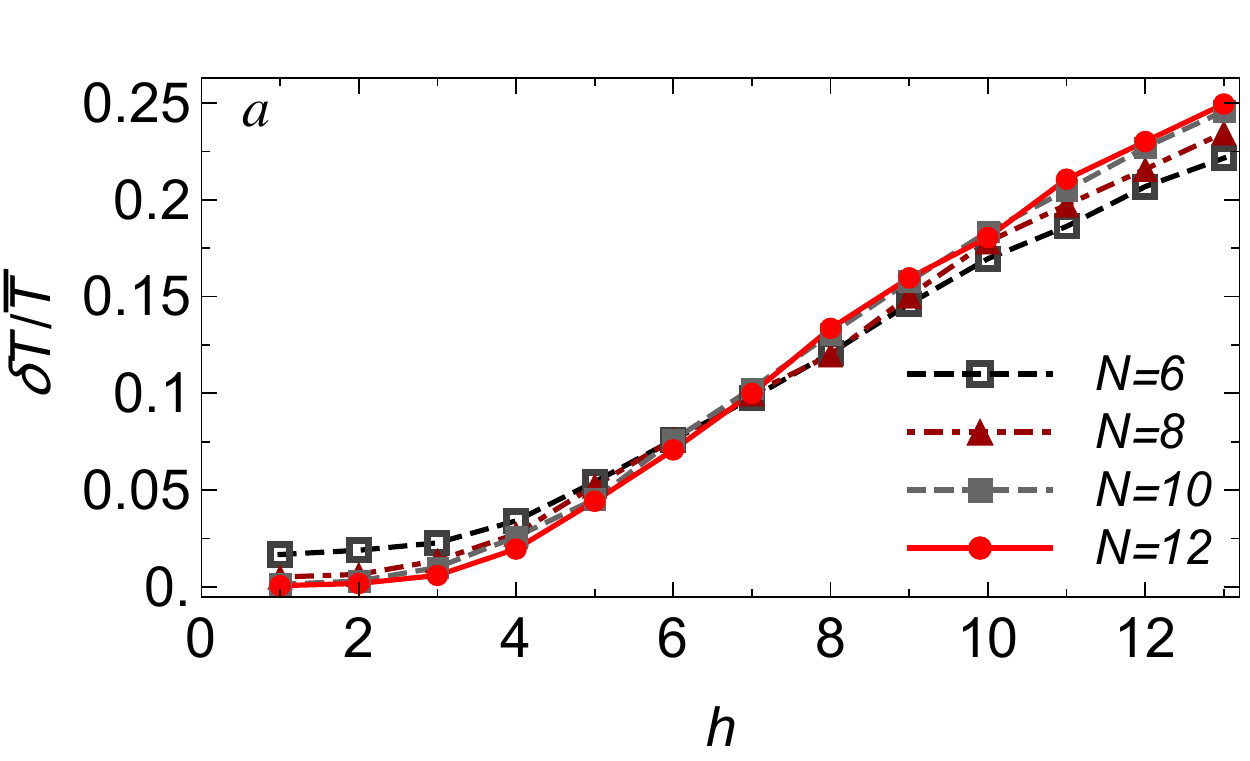}\\
\vspace{-.03\linewidth}
\includegraphics[width=.87\linewidth]{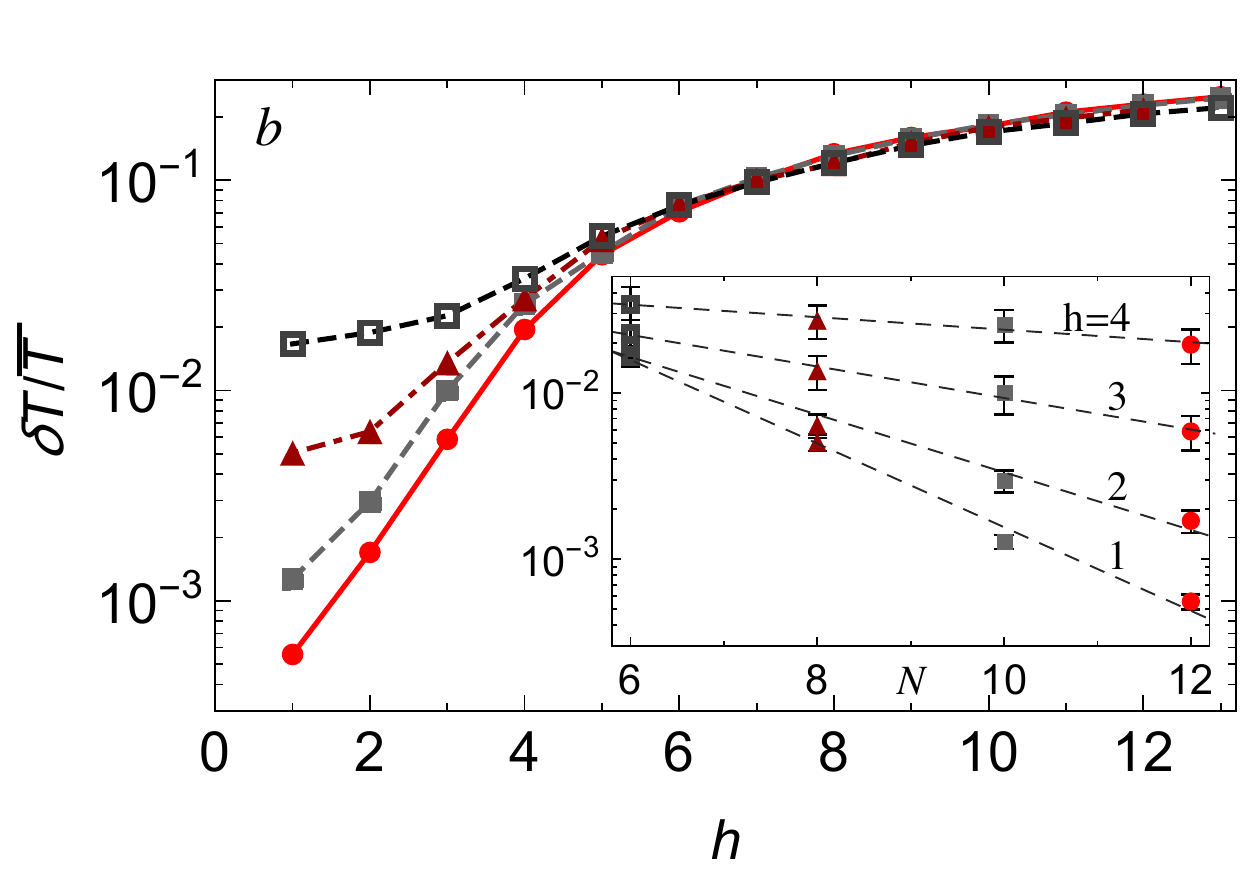}\\
\vspace{-.03\linewidth}
\includegraphics[width=.89\linewidth]{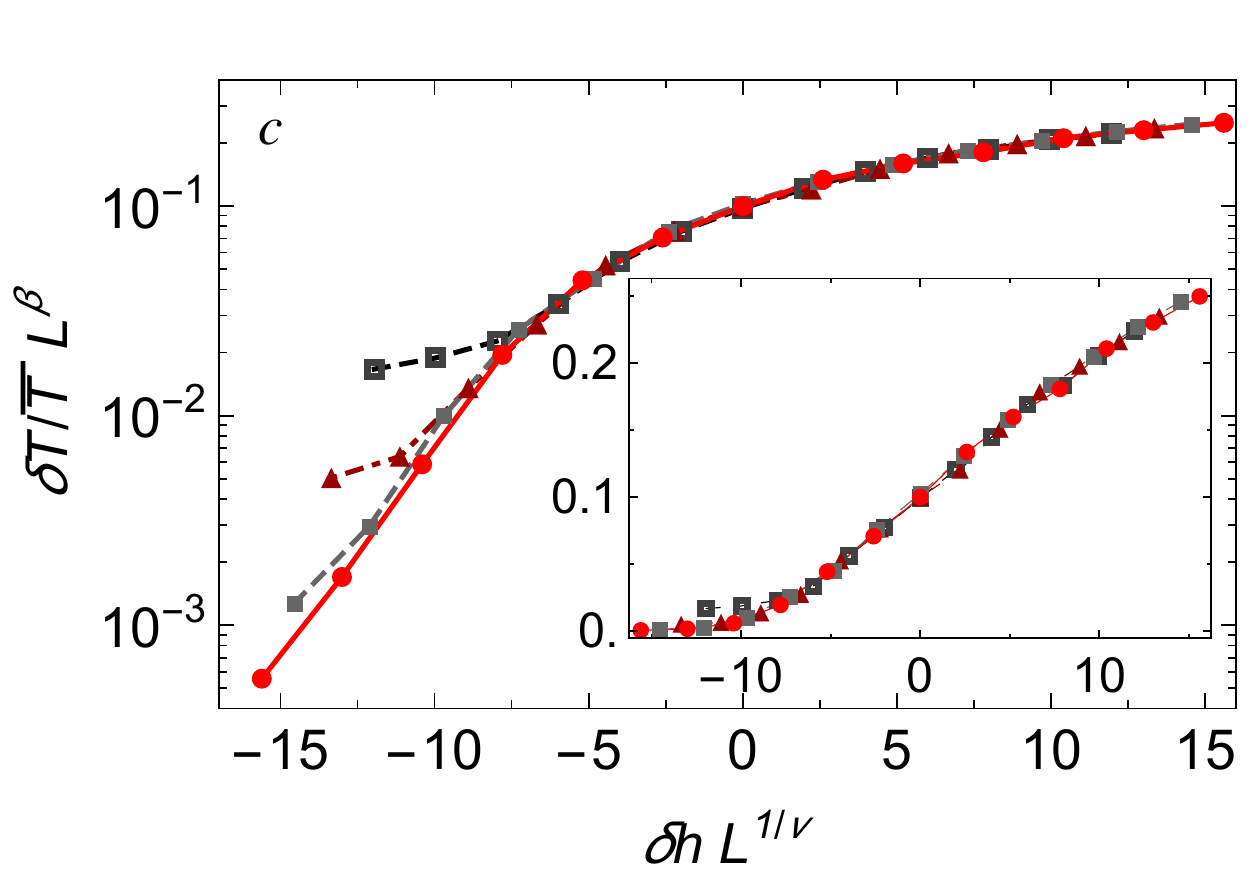}
\caption{\label{FigLocTemp}	
Upper panel: Standard deviation of local temperatures, $\delta T$, as function of disorder strength $h$ for the system sizes $N=6,8,10,12$. 
Middle panel: The logarithmic plot shows that $\delta T$ drops exponentially in system size in the ergodic phase, Eq.~\eqref{ergodicN}. The $N$ dependence for small $h$ is shown in the inset.
Lower panel: A data collapse is obtained by rescaling using $h_c=7$, $\nu=2.6$, $\beta=0$. Inset: same scaling plot  on a linear scale.
Parameters: $T_{p}=10, U=2, (\epsilon_d/\epsilon_p)^2=0.3$.
}
\end{figure}

Fig.~\ref{FigExampleLocTemp}(a) shows the local temperature profile calculated for one disorder configuration and a fixed ratio $(\epsilon_d / \epsilon_{p})^2 = 0.3$ in the limit $\epsilon_d,\epsilon_p\to 0$.  Deep in the ergodic phase the fluctuations of the local temperature are very small while they become large in the MBL regime. 
We propose to use this as an experimental probe of MBL on solids.

We now turn to quantify the magnitude of local temperature fluctuation. For
 each disorder configuration $n$ we determine the deviation of the local temperature $T_{n,i}$ from the average temperature of the chain $T_n\equiv \frac{1}{N} \sum_{i=1}^N T_{n,i}$, that is $\delta T_{n,i}= T_{n,i}- T_n$. The average fluctuation over all sites and disorder configurations is given by $\delta T \equiv \sqrt{\langle{\delta T_{n,i}^2}\rangle}$ and the average temperature is $\bar{T}\equiv \langle T_{n,i}\rangle$. In our numerical results we average over $M=500$ random disorder configurations. 

We expect that in the thermodynamic limit $\delta T=0$  in the ergodic phase, but is nonvanishing in the MBL phase, Fig.~\ref{FigExampleLocTemp}(b). Thus the temperature fluctuation serves as an order parameter of the MBL phase, which is expected to grow with a universal exponent $\alpha$ upon entering the phase
\begin{align}\label{dTinfty}
\lim_{N \to \infty,\epsilon_d, \epsilon_f \to 0} \frac{\delta T}{\bar T} \sim \left\{ 
\begin{array}{ll}
0 & \text{for } h<h_c \\
(h-h_c)^\alpha & \text{for } h>h_c
\end{array}
 \right.
 \end{align}
Here the limit is taken with $\epsilon_d/\epsilon_p=const.$ and $h_c$ is the critical disorder strength characterizing the MBL transition. 

The vanishing of temperature variations for $h<h_c$ follows from the fact that in the thermodynamic limit with $\epsilon_d,\epsilon_p\to 0$ 
 ergodic systems equilibrate to a thermal state characterized by a unique temperature, see Methods for details.
For $h>h_c$, in contrast, an extensive set of local conservation laws prohibits equilibration and we expect a highly non-thermal state arising from the solution of rate equations for which fluctuation dissipation relations are violated, leading to strongly fluctuating local temperatures.

\begin{figure}[t]
\center
\includegraphics[width=.87\linewidth]{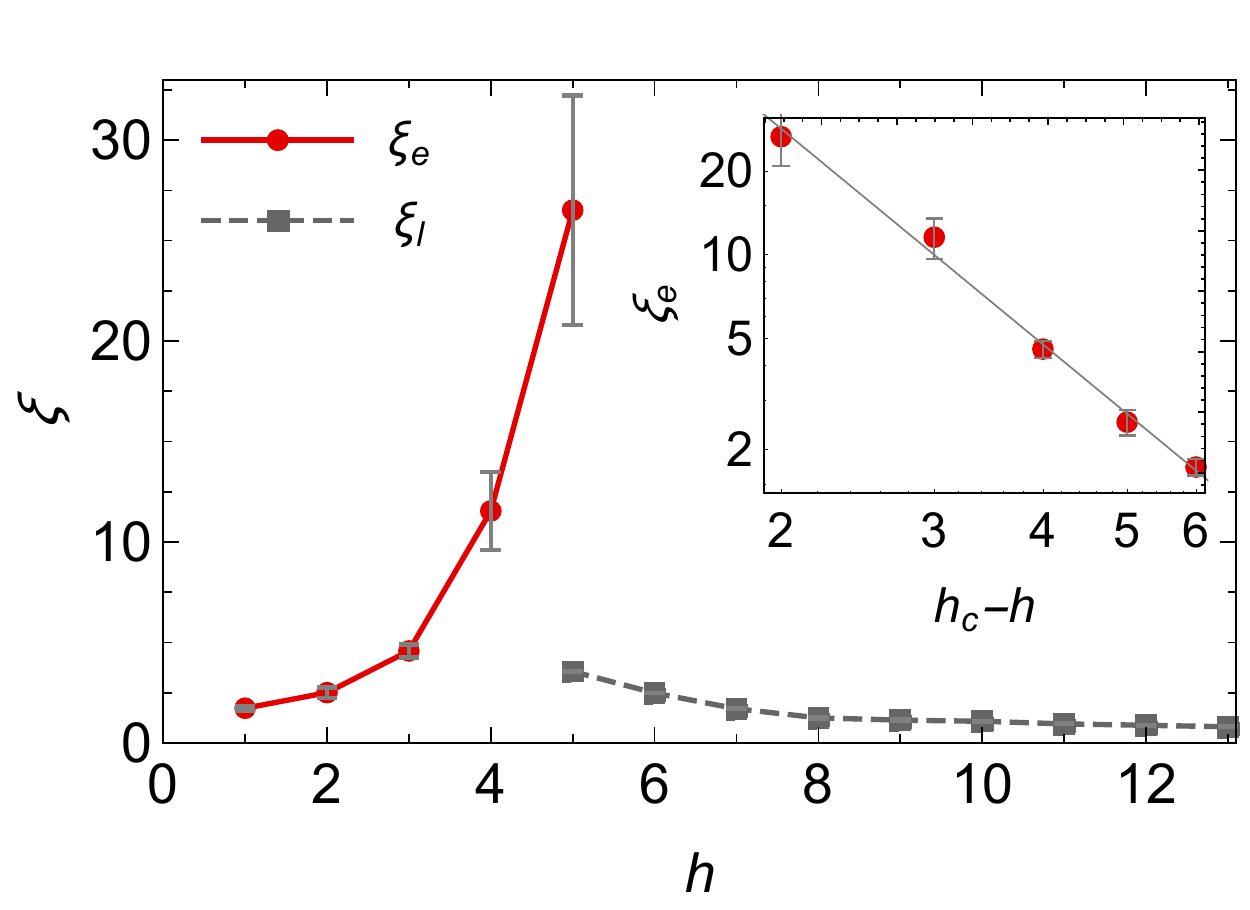}
\caption{\label{FigCorrelLength} 
Correlation length as function of the disorder strength $h$. For small $h$ the correlation length is defined by Eq.~\eqref{ergodicN}, see Fig.~\ref{FigLocTemp}. For large $h$ it is calculated from the spatial correlations of $\delta T$ discussed in Fig.~\ref{FigDTfluct}. Inset: the critical exponent $\nu$ can be extracted from the fit $\xi_e\sim \frac{1}{(h_c-h)^\nu}$. We get $\nu \approx 2.6$ assuming $h_c=7$, see text.
}
\end{figure}

Our numerical calculations are done on a finite size system with up to $12$ sites. In this case the sharp phase transition gives way to a smooth crossover, Fig.~\ref{FigLocTemp}(a). We fix the phonon temperature to $T_{p}=10$, see Methods for a discussion of the dependence on $T_p$.
In the ergodic phase we find that $\delta T$ drops as a function of system size in a manner consistent with exponential dependence, Fig.~\ref{FigLocTemp}(b),
\begin{align}
\frac{\delta T}{\bar T} \sim e^{-L/\xi_{e}(h)}   \label{ergodicN}
\end{align}
The decay length $\xi_e(h)$ grows rapidly on approaching the MBL transition, see Fig.~\ref{FigCorrelLength}, hence we associate it with the correlation length that diverges at the critical point as $\xi_e\sim \frac{1}{(h_c-h)^\nu}$. 

It is instructive to apply a finite size scaling analysis. 
Fig. ~\ref{FigLocTemp}(c) shows a scaling collapse of the data assuming a universal scaling function $\frac{\delta T}{\bar T} = L^{-\beta}f(L^{1/\nu}\delta h)$,  $\delta h=h-h_c$. This scaling function also implies that the order parameter grows as $\frac{\delta T}{\bar T} \sim (h-h_c)^{\beta\nu}$ on crossing the transition. To obtain collapse we assumed $h_c=7$ and fitted $\nu\approx 2.6$, $\beta\approx 0$.   Latter values are consistent also with the fit to $\xi_e(h_c-h)$, see inset of Fig.~\ref{FigCorrelLength}. However, we cannot precisely determine $h_c$ from out data since reasonable data collapse can be obtained within a range of parameters, giving rise to crude estimates  $h_c \approx 6.5 \pm 1$ (consistent with exact diagonalization results, e.g., \onlinecite{luitz15}), $\nu=2.5 \pm 0.5$ and $\beta=0.08\pm 0.08$.
It is interesting that we find a value of $\nu$ consistent with the Chayes-Harris bound \cite{harris74,chandran15}, $\nu>2/d$, 
where $d$ is the spatial dimension. 
This is in marked contrast with results of exact diagonalization studies, which obtain $\nu\approx 1$, e.g. \onlinecite{luitz15}. 
Rather, the result is closer to the renormalization group approaches \cite{vosk15,potter15} which obtain $\nu \approx 3.3$.

\begin{figure}[t]
\center
\includegraphics[width=.87\linewidth]{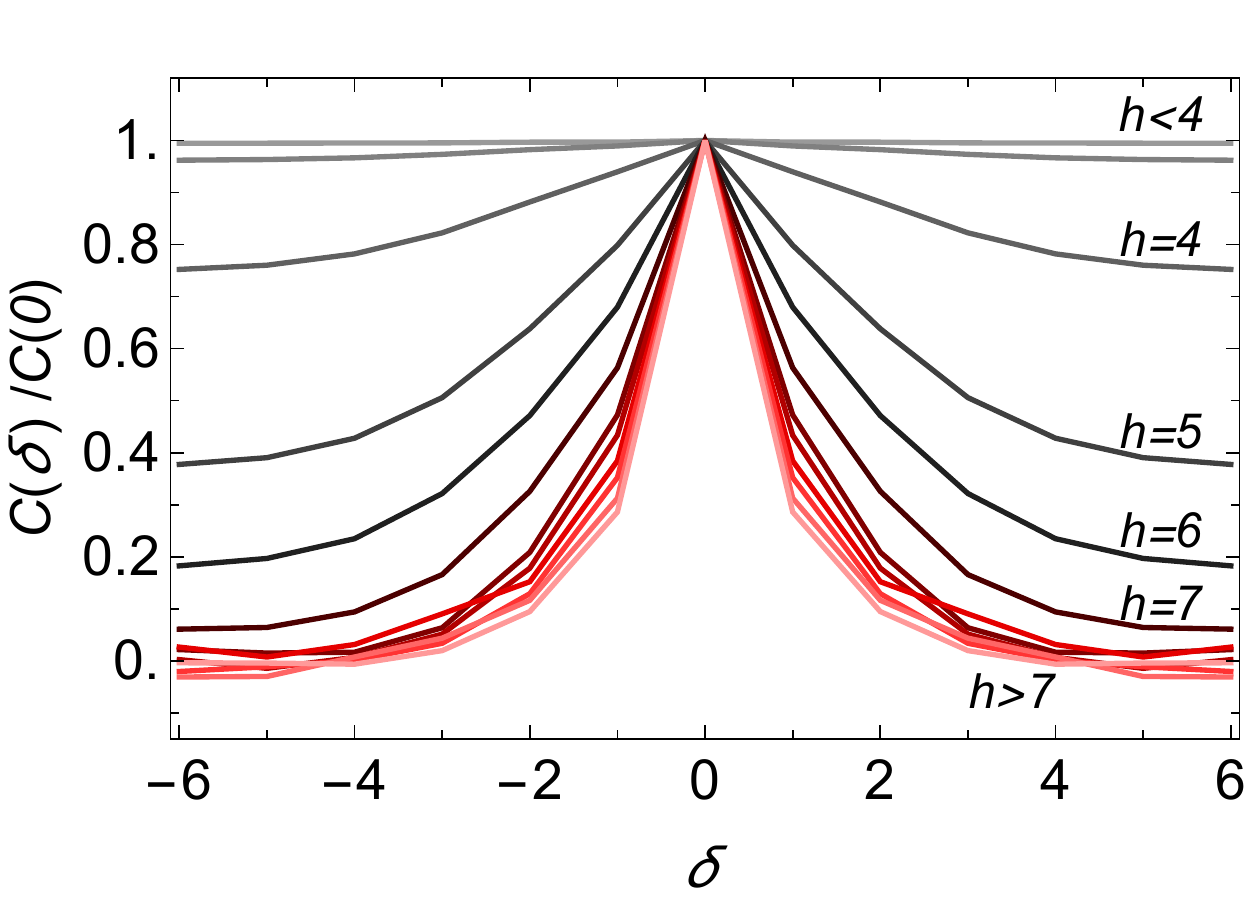}
\caption{\label{FigDTfluct}
Correlation function of the local temperature defined by 
$C(\delta)=\langle (T_{n,i}- \bar{T})(T_{n,i+\delta}-\bar{T}) \rangle$. In the MBL phase temperature fluctuates on a rather short length scale associated with the localization length $\xi_l$. Parameters: $N=12$, $T_{p}=10, U=2, (\epsilon_{d}/\epsilon_p)^2=0.3$.
}
\end{figure}


On the MBL side we expect that $\delta T_j$ fluctuates on short scales associated with the localization length $\xi_l$. The correlation function $C(\delta)=\langle (T_{n,i}- \bar{T})(T_{n,i+\delta}-\bar{T}) \rangle$ is plotted in Fig.~\ref{FigDTfluct}. We determine $\xi_l$  from the fit $C(\delta) \sim  e^{-\delta/\xi_l(h)}+e^{-(N-\delta)/\xi_l(h)}$. 

$\xi_l(h)$ is shown in Fig.~\ref{FigCorrelLength} together with $\xi_e(h)$ obtained from Eq.~\eqref{ergodicN}. $\xi_l(h)$ grows with decreasing disorder, but unlike $\xi_e(h)$ on the ergodic side, it does not seem to diverge at the critical point. This behavior is consistent with other numerical results and  renormalization group approaches \cite{vosk15,potter15}, which also fail to extract a diverging localization length from the behavior of typical physical quantities. Indications of a diverging localization length manifest only when considering special quantities, whose average is sensitive to the appearance of rare thermalizing clusters that ultimately trigger the phase transition to the ergodic phase \cite{vosk15}.

The analysis discussed above is rigorously valid in the limit $\e_p,\e_d\to0$. At finite $\e_p$ and $\e_d$ non-zero fluctuations $\delta T$ are expected also in the ergodic phase. Deep in the ergodic phase, $\delta T$ can be calculated from a straightforward hydrodynamic approach describing the interplay of heat conduction and local heating and cooling by light and phonons, respectively, see Methods. From this analytic approach we obtain in $d$ dimensions deep in the ergodic phase
\begin{align}
&\delta T \sim \frac{ 1 }{\bar \kappa^{d/4}} 
\,\, \left(\frac{\epsilon_d}{ \epsilon_p}\right)^2 \epsilon_p^{d/2}
 \end{align}
 where $\bar \kappa$ is the (average) heat conductivity.
 As expected,  for $\e_p,\e_d\to0$ at fixed ratio $\e_d/\e_p$, $\delta T$ vanishes. Remarkably, the same hydrodynamic approach predicts that $\delta T$ is of order $1$ in the MBL phase, see Methods.
 
As $\delta T$ is finite in the ergodic phase for finite $\epsilon_{p,d}$, the
sharp transition in $\delta T$ expected for $\epsilon_d,\epsilon_p \to 0$, Eq. \eqref{dTinfty},  will be broadened for finite $\epsilon_d,\epsilon_p$, see Fig.~\ref{FigExampleLocTemp}. In an actual solid-state experiment one easily controls $\epsilon_d$ by changing the radiation density but not the strength of phonon coupling. A lowering of temperature does, however, has essentially the same effect as a reduction of $\epsilon_p$ as the ability to cool the system strongly depends on $T_p$, see Methods. By simultaneously lowering temperature and 
irradiation it should be possible to approach systematically the limit $\epsilon_d,\epsilon_p \to 0$.



Finally we discuss possible experimental realizations. Experiments with disordered Indium Oxide films have  shown clear signatures of decoupling between the electron and phonon temperatures, which occurs due to driving the system with voltage together with the weakness of the electron phonon interaction at low temperature \cite{ovadia09}.  Furthermore, precursors of a many-body localization transition at finite T have been reported in the same films \cite{Shahar}.  Hence, we believe this system is a promising testbed for investigating many-body localization in the approach developed in this paper.


Our analysis in this paper mainly focused on the limit of weak coupling to the phonons and the drive, i.e. $\e_p,\e_d\to0$ while keeping $\epsilon_d/\epsilon_p$ constant.  A complete understanding of the driven system at finite $\epsilon_{p,d}$ requires further study. In particular it would be interesting to include the coupling to drive and to phonons within  an effective description of the Griffith phase and the MBL critical point \cite{vosk15}. In addition DMRG can be used to solve for the steady state density matrix of the appropriate Lindblad evolution. 
Such studies could  shed light on how the onset of the order parameter $\delta T/\bar{T}$ broadens into a universal crossover with increasing $\epsilon_p,\epsilon_d$ and thus assist the interpretation of experiments that are necessarily done at finite coupling.

\vspace{0.5cm}
\noindent {\bf Acknowledgement}\\
We acknowledge useful discussions with O. Alberton, M. Knap, and F. Lange. Z.L. and A.R were financially supported by the German Science Foundation under CRC 1238 (project C04) while E.A. acknowledges the ERC synergy grant UQUAM.

\vspace{0.5cm}
\noindent {\bf Author contributions}\\
Z.L implemented the numerics and analyzed the data. All authors jointly
designed the study, interpreted the results and wrote the paper.


\vspace{1cm}
\noindent {\large \bf Methods}

\noindent {\bf Steady state at weak coupling to drive and dissipation.} 
We would like to obtain the steady-state of the driven interacting system in the limit of weak coupling to phonons and light.
Formally we first take the limit $t\to \infty$ and afterwards the limit $\epsilon_p,\epsilon_d \to 0$.
In this sequence of limits the steady state density matrix of the fermionic system is given by
\begin{eqnarray}
\lim_{\epsilon_p,\epsilon_d \to 0} \lim_{t\to\infty}\rho(t)= \sum_n p_n |n\rangle \langle n|,
\end{eqnarray}
where $|n\rangle$ are the exact many-particle eigenstates of $H_f$, $H_f\ket{n} = E_n \ket{n}$. 

The probabilities $p_n$ do, however, depend sensitively on the couplings to phonons and to light, which determine the transition rates $\Gamma_{mn}=\Gamma_{mn}^{p}+\Gamma_{mn}^{d}$ from state $n$ to state $m$. 
The probabilities $p_n$ are computed from  the steady state $d p_n/d t =0$ of the rate equation
\begin{eqnarray}\label{rateequation}
\frac{d}{d t} p_n =  \sum_m \Gamma_{nm} p_m-\Gamma_{mn} p_n.
\end{eqnarray}

For small $\epsilon_p$ and $\epsilon_d$ the transition rates are computed from Fermi's golden rule using the exact eigenstates of $H_f$ obtained from exact diagonalization.

The contribution arising from the couplings to the three-dimensional phonon bath, Eq.~\eqref{EqHph}, is given by
\begin{widetext}
\begin{align}\label{EqLPh} 
\Gamma^{p}_{mn}
= 2\pi \frac{\epsilon_{p}^2}{N} &\sum_{q_x} \int \frac{d^2 \vec q_\perp}{(2 \pi)^2} |\bra{m} H_{q_x} \ket{n}|^2
 \, \Big(n_B(\omega_{\bs q}) \delta(E_m-E_n-\omega_{\bs q})+\big(n_B(\omega_{\bs q})+1\big) \delta(E_n-E_m-\omega_{\bs q})\Big) \notag \\
= 2\pi \frac{\epsilon_{p}^2}{N} &\sum_{q_x} |\bra{m} H_{q_x} \ket{n}|^2
 \, \Big(n_B(E_m-E_n) D_{q_x}(E_m-E_n) +\big(n_B(E_n-E_m)+1\big) D_{q_x}(E_n-E_m)\Big) 
\end{align}
\end{widetext}
where 
\begin{equation}
D_{q_x}(E)= \int \frac{d^2 \vec q_\perp}{(2 \pi)^2} \delta\!\left(E-\omega_{\vec q}\right)=\frac{E}{2\pi v^2} 
\Theta(E - v |q_x|)
\end{equation}
is the phonon density of states for fixed momentum $q_x$ and 
$n_B(E)=1/(e^{E/T_p}-1)$ is the equilibrium Bose distribution at phonon temperature $T_p$.

Similarly, the coupling to light, Eq.~\eqref{EqHd}, simply induces transition rates
\begin{align}
\Gamma_{mn}^d&= 2\pi \frac{\epsilon_{d}^2}{\tilde t}\, |\ave{m|J|n}|^2 
\end{align}
where $J=i \tilde t \sum_{i} (c_{i+1}^\dagger c_i   - c_{i}^\dagger c_{i+1})$ is the current operator.
Note that a $\delta$-correlated vector potential corresponds to the electric field correlation $\langle E_\omega E_{\omega'}\rangle \sim \delta(\omega+\omega') \omega^2$ expected for black-body radiation for $\omega \ll T_d$.
Interestingly, a Lindblad dissipator with the total current as the Lindblad operator would induce the same transitions rates. Such equivalence is a consequence of using a delta-correlated classical field $\langle A(t) A(t') \rangle = 2\pi \delta(t-t')$.

Importantly, the phonon-induced transition rates obey a detailed balance condition, $\Gamma^p_{mn} e^{-E_n/T_p} = \Gamma^p_{nm} e^{-E_m/T_p}$, which guarantees that a thermal state with $p_n^{th} = Z^{-1}e^{-E_n/T_{p}}$ is obtained in absence of the drive (i.e. $\epsilon_d=0$). In contrast, in the absence of coupling to phonons, $\epsilon_p=0$, the system heats up to infinite temperature, $p_n=const.$, as $\Gamma^d_{nm}=\Gamma^d_{mn}$.
When the ratio of $\epsilon_p$ and $\epsilon_d$ is finite, the result is a non thermal state, which violates the equilibrium fluctuation-dissipation relations.

\vspace{0.5cm}
\noindent {\bf Local thermometer.} Local temperatures characterizing the nature of the steady state are determined via an infinitesimal coupling of the system to external bosons, $H_j=\epsilon_l A_j (a+a^\dagger)$. We assume that the thermometer couples via the tunneling term, $A_j= \tilde{t}(c^\dagger_j c_{j+1} + c^\dagger_{j+1} c_{j})$. The local temperature $T_j$ is defined as the temperature of bosons at which  the energy current is zero, $\ave{\dot{H_j}}=0$, 
\begin{align}\label{EqForcePh}
\ave{\dot{H}_j} = 
2\pi \epsilon_{l}^2
\sum_{m,n} &p_n (E_{m} - E_{n})  
\big|\bra{m} A_j \ket{n}\big|^2\\
 \times \Big((&n_B(E_n-E_m,T_j)+1) D^{b}(E_n-E_m)\notag \\
 &+n_B(E_m-E_n,T_j) D^{b}(E_m-E_n) \Big)\notag 
\end{align}
where $n_B(E,T_j)=1/(e^{E/T_j}-1)$ is the Bose distribution at temperature $T_j$. We assume a constant density of bosons, $D^{b}(\omega)= \Theta(\omega)/\tilde{t}$.

\vspace{0.5cm}
\noindent {\bf Breakdown of thermal description  in the MBL phase.}
In the thermodynamic limit with $\epsilon_d,\epsilon_p\to 0$ and constant ratio $\epsilon_d/\epsilon_p$ ergodic systems equilibrate to a thermal state characterized by a unique temperature, $T_{th}$.
$T_{th}$ can be determined from the condition that heating from the irradiation and cooling from the phonon bath compensate each other \cite{lange17},
\begin{eqnarray}\label{Tergodic}
\left \langle  \frac{d H_f}{d t} \right \rangle= \sum_{mn} E_n ( \Gamma_{nm} p^{th}_m-\Gamma_{mn} p^{th}_n) =0,
\end{eqnarray}
with $p^{th}_n=e^{-E_n/T_{th}}/Z$. Note that $T_{th}$ depends on the ratio $\epsilon_d/\epsilon_p$ but not on the absolute value of the two parameters.
The vanishing of $\delta T$ and existence of a homogeneous $T_{th}$ is directly related to the validity of the eigenstate thermalization hypothesis \cite{deutsch91,srednicki94} in combination with the fact that in the thermodynamic limit local observables cannot distinguish the different ensembles, e.g. microcanonical and canonical.

\begin{figure}[t!]
\center
\includegraphics[width=.87\linewidth]{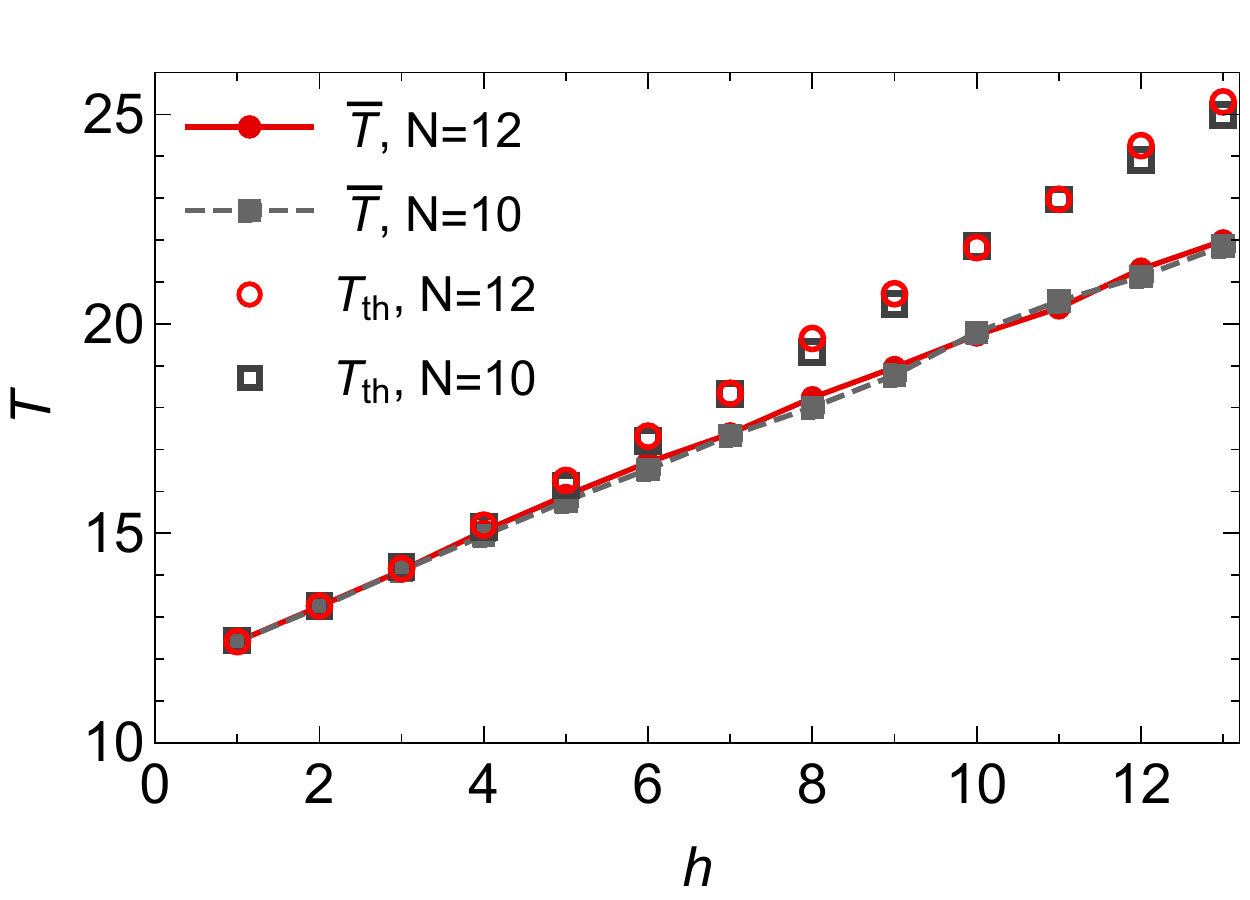}
\caption{\label{FigT}
Comparison of the average temperature 
$\bar{T}=\ave{T_{n,i}}$ 
and the temperature obtained from the thermal ansatz for the density matrix, Eq.~\eqref{Tergodic}, 
$T_{th}=\ave{T_{th}^n}$ averaged over $M=500$ disorder realizations. 
As expected, the thermal ansatz is appropriate only in the ergodic phase, $h \lesssim 6$. Parameters: $N=12, 10$, $T_{p}=10$, $U=2$, $(\epsilon_d/\epsilon_p)^2=0.3$.
}
\end{figure}

In Fig.~\ref{FigT} we compare the average temperature $\bar T$ to the temperature $T_{th}$ obtained from Eq.~\eqref{Tergodic}. As expected, we obtain an excellent agreement of the two temperatures for $h \lesssim 6$, i.e. in the regime where also $\delta T$ is strongly suppressed. Thermal description fails completely in the MBL phase. A similar observation has been made also in Ref.~\onlinecite{deLuca16b}\\

\begin{figure}[t]
\center
\includegraphics[width=.87\linewidth]{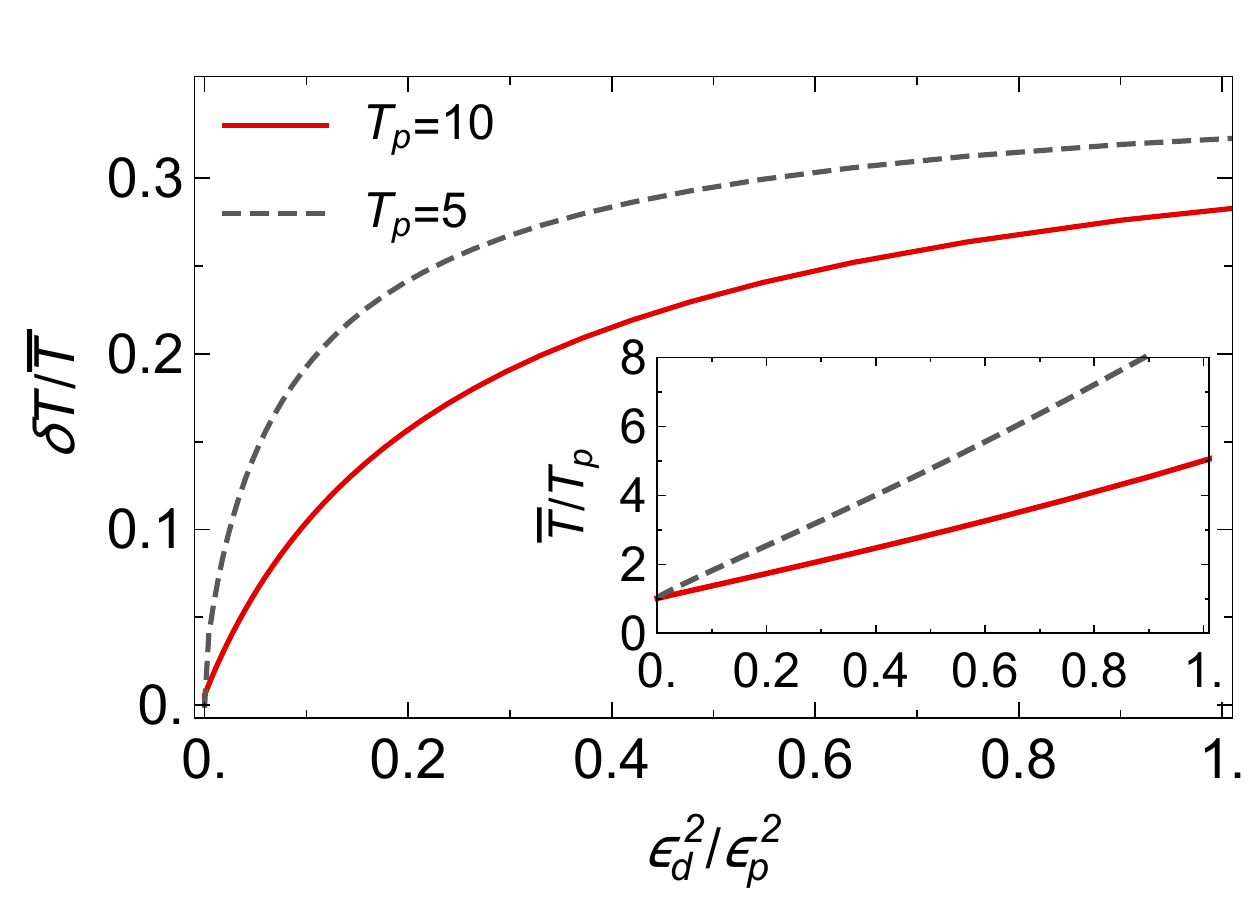}
\caption{\label{FigdTgamma} 
Growth of the fluctuations $\delta T$ as function of $\epsilon_d^2/\epsilon_p^2$ in the MBL phase ($h=10$) for two different phonon temperatures $T_p=5,10$. For lower phonon temperatures much weaker irradiation is needed to obtain the same value of $\delta T$ as at lower $T_p$ phonons are much less efficient in cooling the system. 
}
\end{figure}

\vspace{0.5cm}
\noindent {\bf Dependence of the local temperature fluctuations on the phonon temperature and coupling strength.}
In the results presented in the main text of the paper we have used a phonon temperature $T_p=10$ and a constant ratio $(\epsilon_d/\epsilon_p)^2=0.3$. The qualitative nature of the results does not change if we choose different parameters as long as the phonon temperature and coupling ratio are both finite. However, the absolute magnitude of the local temperature fluctuations in the MBL phase has a strong dependence on these parameters. In particular the fluctuations grow rapidly with increase of the radiation intensity  $\epsilon_d^2$, see Fig.~\ref{FigdTgamma}. The rise in $\delta T/\bar T$ starts linear and saturates when 
$\bar T$ becomes much larger than the phonon temperature. The initial slope of $\delta T/\bar T$ depends strongly on the phonon temperature $T_p$. At lower phonon temperatures the cooling efficiency of the phonons is strongly reduced and weaker irradiation can drive the system out of equilibrium. Therefore lowering the phonon temperature $T_p$  effectively has a similar effect as a reduction of the phonon coupling $\epsilon_p$.

\vspace{0.5cm}
\noindent {\bf Temperature fluctuations in the ergodic phase: hydrodynamic approach.}
In the main text we discussed the limit $\epsilon_d,\epsilon_p\to 0$ for fixed ratio $\epsilon_d/\epsilon_p$. In this limit $\delta T$ vanishes in the ergodic phase in the thermodynamic limit. In an actual experiment, however,  both $\epsilon_d$ and $\epsilon_p$  will be finite and temperature fluctuations are therefore also expected deep in the ergodic phase. 

In this phase, we can use a simple hydrodynamic theory to calculate the fluctuations for small but finite $\epsilon_d$ and $\epsilon_p$. The analytic approach has the advantage that it is not affected by finite size effects and can be formulated for arbitary dimension $d$. It is based on the fact that energy of the fermionic system is the only relevant approximate conservation law. The coupling to light and the phonons act as sources and sinks of energy. They induce fermionic heat currents $\vec j_h(\vec r)=-\kappa(\vec r) \vec \nabla T $ where  $T(\vec r)$ is the local temperature and $\kappa(\vec r)$ the local heat conductivity which depends on $\vec r$ due to disorder.

The continuity equation for the energy, supplemented by sink and source terms, takes the form
\begin{align}\label{continuity}
\partial_t e - \nabla \kappa(\vec r) \nabla T(\vec r) = -  \epsilon_{p}^2 g_{p}(\vec r) (T(\vec r)-T_{p}) + \epsilon_d^2 g_d(\vec r) 
\end{align}
where $e$ is the energy density of the fermionic system. 
$g_p(\vec r)$ and $g_d(\vec r)$ are random functions which describe the local variations of the coupling to phonons and light, respectively. The phonon term proportional to $\epsilon_p^2$ describes the energy transfer to the phonon bath, while the drive term proportional to $\epsilon_d^2$ accounts for heating by the incident radiation.

We are interested in the spatial fluctuations in the steady state, defined by the condition $\partial_t e=0$. Deep in the ergodic phase we can expand around the mean values, 
 $\kappa(\vec r)=\bar \kappa + \delta\kappa(\vec r)$, 
$g_{p}(\vec r)=\bar g_{p} + \delta g_{p}(\vec r)$, $g_{d}(\vec r)=\bar g_{d} + \delta g_{d}(\vec r)$, $T(\vec r)=\bar T + \delta T(\vec r)$ with 
$\bar T=T_{p}+\frac{\epsilon_d^2 \bar g_d}{\epsilon_{p}^2 \bar g_{p}}$. 
The dominant contribution to $\delta T(\vec r)$ is calculated from 
\begin{equation}
(-\bar\kappa \nabla^2 
+\epsilon_{p}^2 \bar g_{p})\delta T(\vec r) 
=\epsilon_d^2 \delta g(\vec r)
\end{equation}
where $\delta g(\vec r)=  \delta g_d(\vec r)-\delta g_{p}(\vec r) \frac{\bar g_d}{\bar g_{p}} $. Note that the heat diffusion equation obtains a mass term due to the coupling to the phonon bath. The corresponding Green function takes in momentum space the form $G(\vec k)=\frac{1}{\bar\kappa k^2 + \epsilon_{p}^2 \bar g_{p}}$.
Using $\delta T(\vec r)=\epsilon_d^2 \int d\vec r' G(\vec r-\vec r')  \delta g(\vec r')$ and assuming short-ranged disorder correlations, $\langle \delta g(\vec r)\delta g(\vec r') \rangle = (\delta g)^2 \delta(\vec r-\vec r')$, we find
\begin{align}
&\delta T =\frac{ c_d |\delta g|  }{\bar \kappa^{d/4} \, \bar g_{p}^{1-d/4}} 
\,\, \left(\frac{\epsilon_d}{ \epsilon_p}\right)^2 \epsilon_p^{d/2}
 \end{align}
where $c_d=1/2,1/ \sqrt{4\pi},1/\sqrt{8 \pi}$ for $d=1,2,3$.

One can also use an equation for energy transport similar to Eq.~\eqref{continuity}  in the MBL phase in the limit when $\epsilon_d$ and the resulting temperature variations are small (instead of the continuous version one needs a lattice version with matrix valued coefficients, e.g. $\kappa_{ij}$). 
The main difference is that for $\epsilon_{d},\epsilon_p =0$ the heat conductivity $\kappa$ vanishes.
Heat transport in the MBL phase is only possible due to transitions between localized levels induced by the coupling to phonons or the drive. As a consequence in the MBL phase the thermal conductivity itself depends on the coupling to phonons and to the drive as
\begin{eqnarray}
\kappa(\vec r) \sim O(\epsilon_d^2,\epsilon_p^2).
\end{eqnarray}
In this case all terms in Eq.~\eqref{continuity} become proportional to either $\epsilon_d^2$ or $\epsilon_p^2$. For this reason
$\delta T$ remains of $O(1)$ and depends only on the ratio $\epsilon_d/\epsilon_p$ and not on the absolute value of the two parameters.



\end{document}